\documentclass[letter, traditabstract]{aa}
\usepackage{graphicx}
\usepackage{txfonts}
\begin{document}
   \title{Period  doubling and non-linear resonance in the black hole
          candidate IGR J17091-3624 ?}

   \author{P.~Rebusco \inst{1} 
           \and
           P.~Moskalik  \inst{2}
           \and
           W.~Klu{\'z}niak \inst{2}
           \and
           M.A.~Abramowicz  \inst{3,2,4} 
           \fnmsep\thanks{A visitor at Charleston College, SC USA.}
          }

   \institute{Experimental Study Group, Massachusetts Institute of Technology,
           77 Massachusetts Avenue, 02139 Cambridge (MA), USA\\
           \email{pao@space.mit.edu}
           \and
           Copernicus Astronomical Center, ul. Bartycka 18, PL-00-716 Warszawa,
           Poland\\
           \email{wlodek@camk.edu.pl}, \email{pam@camk.edu.pl}
           \and
           Department of Physics, University of Gothenburg,
           SE-412-96 G{\"o}teborg, Sweden\\
           \email{marek.abramowicz@physics.gu.se}
           \and
           Institute of Physics, Faculty of Philosophy and Science,
           Silesian University in Opava,
           Bezru{\v c}ovo n{\'a}m. 13, CZ-746-01 Opava, Czech Republic
          }

   \date{Received January 25, 2012; accepted March 10, 2012}

\abstract {The two high frequency quasi periodic oscillations (HFQPOs)
  recently reported in the black hole candidate IGR~J17091-3624 by
  Altamirano and Belloni (2012) are in a 5:2 frequency ratio (164 Hz
  to 66 Hz). This ratio is strongly suggestive of period doubling and
  nonlinear resonance analogous to phenomena known in RV~Tauri-type
  pulsating stars (and recently discovered also in oscillations of
  RR~Lyrae-type and of BL~Herculis-type variables). An interpretation
  of the frequency ratio in terms of nonlinear interactions and a
  comparison with the HFQPOs reported in GRS 1915+105 may imply a mass of
  about 6 solar masses for the black hole in~IGR~J17091-3624.}



\maketitle
%
\section{Introduction}

The source IGR J17091-3624 has received some notoriety as a
possibly low mass ($M\sim 3  M_\odot$) black hole candidate
(Altamirano et al. 2011). The suspected low mass value has been
questioned by Altamirano and Belloni (2012) who discovered two
high frequency quasi periodic oscillations (HFQPOs) with
frequencies 66 Hz (at $8.5$ sigma) and 164 Hz (at $4.5$ sigma).
We point out that the two frequencies are in a 5:2 ratio.

If the underlying fundamental mode of oscillations corresponds to that
responsible for the pair of HFQPOs in a 5:3 ratio previously observed
in another black hole transient (GRS 1915+105), then the black hole in IGR
J17091-3624 would have a run-of-the-mill mass, with a value comparable
to that of XTE J1550-564
(which is reported to have $9.10 \pm 0.61 M_\odot$, Orosz et al. 2011).
The 5:2 ratio further strengthens the case
for non-linear resonance of accretion disk oscillations as the source
of HFQPOs.

The resonance interpretation was first suggested by Klu\'zniak and
Abramowicz (2001), who argued that the QPO phenomenon
(van der Klis 2000) cannot be of
a kinematic origin but must be due to accretion disk oscillations,
and that the twin-peak HFQPOs are due to resonances between
particular modes of disk oscillations.
These authors (Abramowicz and Klu{\'z}niak, 2001)
also noticed  that the 450 Hz frequency in the X-ray flux of
the black hole candidate GRO J1655-40 discovered by Strohmayer
(2000) was in a 3:2 ratio to the previously known 300 Hz frequency
of a HFQPO in the same source.

Subsequently, Klu\'zniak and Abramowicz (2002) noted that {\it ``a 2:3
  ratio [is] in agreement with the 300 and 450 Hz frequencies reported
  in GRO J1655-40 and the 184 and 276 Hz frequencies reported in XTE
  J1550-564. [...] a 5:3 ratio [is] in agreement with the 69.2 and
  41.5 Hz frequencies reported in GRS 1915+105''}, and interpreted the
observations in terms of parametric resonance of epicyclic
oscillations in General Relativity (GR).

Specifically, a rotation supported (slender) fluid torus about a black
hole has modes of oscillation with eigenfrequencies corresponding to
the radial and vertical epicyclic frequencies in the Kerr metric. In
parametric resonance these eigenfrequencies are in a 3:2 ratio, and
the first overtone of the vertical mode adds a third frequency for a
sequence of ratios of 5:3:2. A more general discussion of possible 3:2
orbital resonances, and the resulting spin estimates for microquasar
black holes is given in T{\"o}r{\"o}k et al. (2005).  The epicyclic modes are
always among fundamental modes of oscillation of fluid disks and tori
around black holes and neutron stars (see, e.g., Blaes et al. 2007;
Blaes, Arras and Fragile 2006, and references quoted there). However,
as higher frequencies in a 3:2 ratio were subsequently reported in GRS
J1915+105 (\cite{mcr04,rem06}),
the identification of the 69.2 Hz oscillation with the overtone of
parametric epicyclic resonance was abandoned (\cite{kato}).

In this {\it Letter}, we would like to present a unified scheme for the
HFQPOs in IGR J17091-3624 and GRS J1915+105, and their
frequency ratios, which is  not necessarily tied to any
specific theoretical model of accretion disk oscillations.

\section{Period doubling and half-integer frequencies}

We would like to suggest a simple interpretation of the 5:3 and 5:2
ratios of the reported frequencies in black hole transients that has a direct
analogue in phenomena observed in certain pulsating stars. One
reason for framing the discussion in this way is that although
accretion disks, just like stars, are fluid bodies in (or near)
hydrostatic equilibrium, the internal physical properties (such as the
distribution of pressure or density) of accretion disks are not well
known and this introduces an additional uncertainty in modeling their
oscillations.

As already pointed out in the context of HFQPOs (Abramowicz \&
Klu\'zniak, 2003), a subharmonic (at half the fundamental
frequency) is a hallmark of non-linear interactions.
The presence of subharmonics is a frequency domain manifestation
of period doubling. In the time domain, it corresponds to
oscillations with twice the original period and with alternating
cycles of higher and lower maxima and  deeper and shallower minima. 
Such behavior is
well known in some oscillating stars. It has been observed for
decades in the RV~Tauri-type variables. In fact, such alternations
of the lightcurve are the very definition of this class of
pulsators. A frequency analysis of RV~Tauri-type stars reveals the
presence of subharmonics at 1/2 and at 3/2 of the fundamental oscillation
frequency, $f$ (Pollard et al. 1996; Koll\'ath et al. 1998). Similar
subharmonics are also occasionally detected in the W~Virginis-type
stars (e.g., Templeton \& Henden 2007), which are closely related
to the RV~Tauri-types variables.
In the last two years, period doubling and subharmonics have been
discovered in two other classes of pulsators: in the RR~Lyrae-type
stars (Szab\'o et al. 2010) and in the BL~Herculis-type stars
(Smolec et al. 2012). 

In most cases, more than one subharmonic is present. This is a
result of the strong nonlinearity of oscillations, which generates
harmonics of the main frequency (at $2f$, $3f$, etc.), but also odd
multiples of the subharmonic (at $3/2\,f$, $5/2\,f$, etc.). The most
extreme example of this behavior is seen in the prototypical variable
RR~Lyrae, where a complete sequence of subharmonic multiples from
$1/2\,f$ up to $27/2\,f$ has been detected (Kolenberg et
al. 2011). However, BL Her type stars exhibit only a few odd multiples
of the subharmonic, in one case the observed sequence being $1/2\,f$,
$5/2\,f$, $7/2\,f$, $9/2\,f$.  It is not understood why $3/2\,f$ fails
to show up, but that is often the case in pulsating stars---some, but
not all, of the predicted (allowed) frequencies are present.


The period doubling phenomenon in oscillating stars is well
understood and is reproduced by state-of-the art numerical
hydrodynamical models. For the BL~Herculis-type stars it was
actually predicted 20 years prior to its discovery (Buchler \&
Moskalik 1992). 
By careful analysis of the models, the origin of period doubling in
oscillating stars has been traced to the occurence of a half-integer
resonance of the type $f_k/f \simeq n+1/2$ between the fundamental
mode and another oscillation mode of frequency $f_k$
(Moskalik \& Buchler 1990).

 In the BL~Herculis-type stars period doubling and the concomitant
 subharmonic frequencies are induced by the 3:2 resonance
 (Buchler \& Moskalik 1992; Smolec et al. 2012), 
 in the RR~Lyrae-type stars by the 9:2 resonance
 (Koll\'ath, Moln\'ar \& Szab\'o 2011), while in the case of the
 W~Virginis and RV~Tauri-type variables by the 5:2
 resonance (Moskalik \& Buchler 1990) between the fundamental mode and
 a radial overtone. We note that resonances in such ratios are
 possible also in accretion disks.  Hor\'ak (2004) and Rebusco (2004)
 independently considered geodesic perturbations and indeed found the
 possibility for a 3:2 ratio and other $m$:$n$ ratios between vertical
 and radial epicyclic oscillations. For plane-symmetric resonances $n$
 is even, and the 3rd order resonance occurs when $m/n = 3/2$, the 5th
 order resonance occurs when $m/n = 5/2$, and so on, with the higher
 order resonances being less probable.

Taking our cue from pulsating stars, we could assume that the two black hole
systems under discussion exhibit various odd multiples of the
subharmonic $f/2$ of their respective fundamental frequencies, $f$.
In particular, in GRS 1915+105 one could assume that the two HFQPO
frequencies that are in a 5:3 ratio correspond to $5/2\,f$ and
$3/2\,f$, while in IGR J17091-3624 one observes the fundamental $f$,
and $5/2\,f$.
Assuming that in both cases the fundamental frequency corresponds
to the same mode of oscillations, one can compare the source
masses (assuming similar values of the dimensionless spin
parameter $a_*$), as in GR the mode frequencies are inversely
proportional to the black hole mass. In this scheme, the
fundamental in GRS 1915+105 would be 2/3 of the lower frequency
in the observed pair 69.2 Hz and 41.5 Hz, i.e., about 28 Hz---we
note that a 27 Hz frequency was also detected in this source
by Belloni et al. (2001).
The lower frequency reported in IGR J17091-3624, 66 Hz, is about
twice the fundamental frequency inferred for GRS 1915+105, and the
same ratio (2.4) is, of course, obtained from the higher reported
frequencies, 164 Hz and 69 Hz. This ratio would imply that the
black hole mass of IGR J17091-3624 is about half the mass of the
black hole in GRS 1915+105, which was reported to be $14\pm
4M_\odot$ \cite{gre01}. 
Thus, the black hole in IGR J17091-3624
should have a mass of about $6 M_\odot$ (modulo the unknown
spin).

So far, we have been referring to period doubling in the sense of
  its purely observational consequences of alternating higher and
  lower maxima and deeper and shallower minima in the light curve or,
  equivalently, the presence of subharmonics in the frequency
  domain. However, the term `period doubling' is used to denote a
  bifurcation in a dynamical system, a transition from one periodic
  state to another, period doubled one. In some cases this is
  followed by a cascade of doublings, which can lead to chaos (and
  quasi-periodicity).  In the case of pulsating stars, numerical
  models display the period doubling bifurcation. It can also be
  demonstrated that the original periodic solution looses its
  stability at the onset of bifurcation. Such an analysis has not yet
  been performed for accretion disk oscillations, so our suggestion of
  period doubling in the two black hole systems does not have the strong
  theoretical underpinning that period doubling has in BL Her, RR Lyr
  and RV Tau type stars. One has to be open to the possibility that
  the 5:2 and 5:3 frequency ratios in IGR J17091-3624 and GRO
  J1915+105 may be a manifestation of a direct resonance between two
  modes of oscillation (Klu\'zniak \& Abramowicz 2001; Abramowicz \&
  Klu\'zniak 2002; T{\"o}r{\"o}k et al 2005), which does not necessarily
  lead to period doubling. For instance, a 3:2 coupling will lead
  to period doubling if the higher frequency mode is resonantly driven
  by the lower frequency mode (Moskalik \& Buchler 1990), but not vice
  versa.

%

\section{Properties of stellar and disk oscillators}
We have referred to three classes of stellar pulsators, RV Tau
type stars, RR Lyrae type stars and BL Herculis type stars, 
in which the growth rates of the oscillations (e.g., thousands of days
in RR Lyrae) are much shorter than other characteristic timescales,
such as the thermal timescale ($10^8$ years in RR Lyrae), but much
longer than their periods. The amplitudes of the oscillation have
reached saturation. Thus the oscillations are often essentially
strictly periodic (or multiperiodic). The subharmonic oscillations
typically have amplitudes much lower than the harmonics, but they are
not always periodic: in RR Lyrae the main oscillation is amplitude
modulated and period doubling is present in some stretches of data,
but not in others (Szab\'o et al. 2010, Kolenberg et al. 2011),
thus leading to a finite width of the subharmonics in the frequency
 domain. In RV Tauri type stars, the subharmonics also have a finite width.

The excitation and damping mechanism of the accretion disk
oscillations (Kato 2001, Wagoner 2002) is without doubt very different
from that in the stellar pulsators discussed above. When the QPOs are
present the disk is probably in a very turbulent state and varies on
all timescales, including scales which are not very much longer than
the dynamical timescale. This makes modeling the oscillations a
challenging task. If we were to search for similar behavior in stellar
systems, we would look to red giants and solar type stars, where
short-lived oscillations are stochastically excited by turbulent
convection (e.g., De Ridder et al. 2009,
Goldreich \& Keely 1977, Houdek 2010).  
Period doubling has not (yet?) been reported in those
stars.

With only two HFQPOs recently discovered in J17091-3624 it is too
early to be confident about the presence of a subharmonic series
similar to the one observed in the  pulsating stars, even though the
5/2 subharmonic is much weaker than the presumed fundamental at 66 Hz.
While we have reinterpreted the GRS 1915+105 QPO frequencies in
light of the newly discovered ones in J17091-3624, it is important to
note that simple harmonics of the presumed fundamental at 27 Hz have
not yet been reported in that source.

The black hole HFQPOs can only be extracted from the data upon
integration of the X-ray signal for thousands of oscillation periods, the
resulting average Lorentzian profiles fitted to the data have widths
comparable to the frequency (quality factor, $Q$, of a few)---it is not
clear whether this reflects an intrinsically low coherence of black hole disk
oscillations or whether the frequencies shift during the integration
time. We note that in neutron stars, where the signal is stronger, at
least some HFQPOs have been shown to have a larger coherence, with a
quality factor of $Q\approx200$ (Barret et al. 2005, Mukherjee \&
Bhattacharya 2011).   These values are not so different from those in
oscillating red giants: in $\epsilon$ Oph oscillation modes have periods
of $\sim6$ hours, with a lifetime of possibly $3$ days,
for a quality factor value
of $Q\approx10$ (Barban et al. 2007), and in another source, 
CoRoT 101034881 the corresponding numbers are  about 6 hours,
and $50$ days, for a $Q\approx200$ (De Ridder et al. 2009). 
In the Sun, the 5 minute oscillations have 
 a  $Q\sim10^3$,
e.g., the 4.9 min period oscillation has a lifetime of 0.7 days
for a $Q=270$, while the 8.2 min period corresponds to a lifetime
of 17.1 days, for a $Q=3000$ (Libbrecht 1988).

%
\section{Discussion and Conclusions}

\noindent In this Letter we suggest a possible interpretation of
the recently observed HFQPOs in the black hole candidate IGR
J17091-3624. By analogy with stellar oscillations, we interpret
HFQPOs in this source (as well as in GRS 1915+105) in terms of
non-linear resonances of oscillating modes of the accretion disk.
The present work is not concerned with a specific model for the
origin of such oscillations: it simply illustrates universal
features of non-linear interactions that match observations, and
lead to  a reasonable mass estimate (of about $6 M_\odot$) for IGR
J17091-3624 on the assumption that the 66 Hz oscillation
corresponds to the 27~Hz oscilation in GRS 1915+105.
The common denominator between accretion disks and
stars is that they are fluid bodies near hydrostatic
equilibrium (although with different accuracy: RR Lyrae can be
considered stationary on a timescale of $10^8$ years, while the state of
the disk in a black hole transient can change in minutes).
The fact that in one case GR effects are important while in the
other they are not, may be relevant in the discussion of the
nature of the fluid oscillations, but it does not affect the
simple period doubling phenomenon.

  If confirmed, period doubling in black hole QPOs could be the
  first such detection in a stochastically driven astrophysical
  system. A more detailed investigation of the full spectrum of QPOs
  in IGR J17091-3624 is eagerly awaited, as it could reveal a sequence
  of harmonics and subharmonics with amplitude structure expected in a
  period doubling system or, alternatively, suggesting the presence of
  a direct high-order resonance.


%
\begin{acknowledgements}
This work was supported by Polish grants NN203381436 and NCN
2011/01/B/ST9/05439. MAA has been supported by the Czech grant MSM 4781305903.
WK thanks the hospitality of the Aspen
Center for Physics, where a part of this work was completed.
\end{acknowledgements}


\end{document}